\UseRawInputEncoding
\documentclass[final,onefignum,onetabnum]{siuro250211}

\usepackage{lipsum}
\usepackage{amsfonts}
\usepackage{graphicx}
\usepackage{epstopdf}
\usepackage{algorithmic}
\usepackage[numbers]{natbib}
\usepackage{makecell}
\usepackage{float}

\ifpdf
  \DeclareGraphicsExtensions{.eps,.pdf,.png,.jpg}
\else
  \DeclareGraphicsExtensions{.eps}
\fi

\usepackage{enumitem}
\setlist[enumerate]{leftmargin=.5in}
\setlist[itemize]{leftmargin=.5in}

\newsiamremark{remark}{Remark}
\newsiamremark{hypothesis}{Hypothesis}
\crefname{hypothesis}{Hypothesis}{Hypotheses}
\newsiamthm{claim}{Claim}
\newsiamremark{fact}{Fact}
\crefname{fact}{Fact}{Facts}

\headers{Developing a Sequential Deep Learning Pipeline to Model Alaskan Permafrost Thaw Under Climate Change}{A. Rahaman}

\title{Developing a Sequential Deep Learning Pipeline to Model Alaskan Permafrost Thaw Under Climate Change}

\author{Addina Rahaman\thanks{The City College of New York, NY
  (\email{arahama003@citymail.cuny.edu}).}
}

\dedication{\small\textit{Project advisor: María Isabel Sánchez-Muñiz}}

\usepackage{amsopn}

\ifpdf
\hypersetup{
  pdftitle={Developing a Sequential Deep Learning Pipeline to Model Alaskan Permafrost Thaw Under Climate Change},
  pdfauthor={A. Rahaman}
}

\begin{document}
\raggedbottom

\maketitle

% REQUIRED
\begin{abstract}
Changing climate conditions threaten the natural permafrost thaw-freeze cycle, resulting in elevated year-round soil temperatures above 0\textdegree C. In Alaska, the warming of the topmost permafrost layer, regarded as the active layer, signals increased greenhouse gas release due to high carbon storage. Therefore, accurate soil temperature predictions are essential for risk mitigation and stability assessment; however, many existing approaches neglect the multitude of variables that influence soil thermal dynamics. This study presents a proof-of-concept implementation of a latitude-based sequential deep learning pipeline to model yearly soil temperatures across multiple depths. The pipeline employs dynamic reanalysis feature data from the ERA5-Land dataset under ECMWF, additional static geologic and lithological features, sliding-window sequences for seasonal context, a derived scenario signal feature for long-term climate forcing, and latitude band embeddings for spatial sensitivity. Five established deep learning models were tested: a Temporal Convolutional Network (TCN), a Transformer, a 1-Dimensional Convolutional Long-Short Term Memory (Conv1DLSTM), a Gated-Recurrent Unit (GRU), and a Bidirectional Long-Short Term Memory (BiLSTM). Performance results presented a solid understanding of latitudinal and depth-wise temperature discrepancies and sound seasonal pattern recognition across all models, suggesting this pipeline can be extended to numerous relevant variables outside of the scope of this study. In particular, the GRU presented the strongest insight of sequential temperature patterns. Representative concentration pathway (RCP) data from the IPSL-CM5A-MR model were bias-corrected using quantile mapping to align with ERA5-Land data, enabling model recognition of sinusoidal temperature trends; however, scenario data constraints in IPSL-CM5A-MR limited divergence between scenarios. Further experiments highlight the contribution of the scenario signal feature, temporal feature derivations, and accumulated snowfall to model predictions. The overall result of this study establishes an end-to-end framework for adopting deep learning in active layer temperature modeling, offering seasonal, spatial, and vertical temperature context without intrinsic restrictions on feature selection.

\end{abstract}

\section{Introduction}
Permafrost is formally defined as a perennially frozen soil, rock, or ice layer \citep{Ran2022, Schuur2008}. Approximately 15\% of the surface of the planet, 22\% of the Northern Hemisphere, and 85\% of Alaska, the largest state in the US by land area, is covered with permafrost \cite{WANG2023165709}. The risks of permafrost thaw from human activities and changing climate conditions continue to be an increasing contemporary threat due to increased carbon output. Permafrost degradation undergoes a positive feedback loop as carbon outputs compound over time, guaranteeing the inevitability of prolonged thawing of the permafrost, deterioration of the infrastructure and disruption of local ecology and hydrology \citep{Jafarov2012, Muniz2025}. Permafrost projection models are critical in forecasting the state of permafrost climate scenarios and in mitigating potential hazards. The surface layer of permafrost in particular is paramount to permafrost modeling. Commonly referred to as the active layer, this layer is defined by its seasonal thawing and refreezing behavior. Observations within the active layer, such as \textbf{active layer thickness (ALT)} and its thermal state, are common proxies for measuring the stability of the subsurface permafrost, the depth of thaw and the potential for carbon release \cite{Chen2024}; they are hence valuable for developing insightful permafrost projections. 

Multiple forms of active layer models have been developed, such as physics-based models, conceptual climate models, and \textbf{Machine Learning (ML)} models \citep{Chance2024, Jafarov2012, Muniz2025}. Physical models such as the Geophysical Institute Permafrost Laboratory (GIPL) model simulate heat transfer through the soil using the heat equation, which consists of numerous parameters involving energy balance, thermal properties of snow, soil moisture, air temperature and vegetation cover \cite{Jafarov2012}. However, traditional models often lack latitude-based projections, limiting their ability to simulate spatial climate scenarios. Conceptual models, on the other hand, can be translated into latitudinal modeling. The Energy Flow Model of S\'{a}nchez-Mu\~{n}iz \cite{Muniz2025} employs an energy-based framework to capture the temperature dynamics of the soil with variables such as the zero-curtain effect and the energy of fusion at play. Part of the motivation for exploring this approach is to better understand its connection to planetary energy balance and to investigate how other models might be coupled with energy balance frameworks, such as the Budyko model \cite{Widiasih2013}, which outputs latitudinal temperature. 

Machine learning is a newer approach, which uses statistical analysis and linear algebra to model multi-variable scenarios involving geospatial and temporal variables in active layer and climate behavior. It bridges the gap between conceptual modeling and empirical observations. However, most machine learning studies on permafrost have largely utilized simple regression-based models such as XGBoost and ensemble models such as RandomForest, which treat time points as flat, independent feature vectors lacking temporal or latitudinal context, thus failing in seasonal pattern detection capabilities. Chen et al.\cite{Chen2024} applied machine learning to predict ALT using environmental variables and geospatial interpolation. However, this approach lacked an understanding of seasonal climate dynamics and does not incorporate energy balance components. On the other hand, Chance et al. \cite{Chance2024} trained numerous regression-based and tree-based models, along with a Multilayer Perceptron (MLP) neural network, on individual seasonal datasets to predict season-based permafrost soil temperatures. These models performed well on short-term vector data but failed to capture observed permafrost behaviors, such as the zero-curtain effect, which reflects decades of thermal and hydrological trends \cite{Anisimov2006}. Moreover, current machine learning methods for permafrost are not latitude-aware despite latitude being essential for modeling active layer behavior \cite{Schuur2008} and they overlook the compound effects of increasing carbon outputs. \textbf{Deep learning (DL)} approaches effectively address these limitations.

%These models performed well on short-term vector data; however they do not match observed permafrost behaviors (such as zero-curtain effect), which remembers decades of thermal and hydrological trends \cite{Asiminov2006}.  Additionally, current permafrost machine learning methods are not latitude-aware, despite latitude being essential in modeling active layer behavior \cite{Schuur2008}, and also fail to account for compound effects of increasing carbon outputs. These limitations are effectively addressed by deep learning (DL) approaches. 

Deep learning (DL) methods excel in climate analysis by retaining long-term spatiotemporal patterns, filtering noise, and capturing complex dependencies. Unlike classical machine learning, which treats time points as independent vectors, DL uses neural networks to model seasonal, latitudinal, and geospatial climate patterns shaped by long-range interactions. This is crucial for understanding processes such as positive carbon feedback loops that drive soil temperature changes. DL also uses dense vector representations of categorical data known as \textbf{embeddings}, allowing the model to distinguish permafrost behavior across latitudes. Most importantly, DL models are highly flexible: as long as relevant numerical or categorical data can be provided, they can identify key features without extensive domain-specific preprocessing.

%Deep learning methods, unlike its classical machine learning counterparts, would be most beneficial in climate analysis due to its ability to retain long-term spatiotemporal patterns by processing sequences of data and disposing of noise. Inspired by neuron connections in the human brain, DL would be advantageous when learning seasonal, latitudinal, and geospatial climate patterns, which are influenced by long-range dependencies rather than independent flat vectors. Such long-range dependencies can provide valuable insight towards the long-term effect of positive carbon feedback loops to soil temperature. DL can also represent categorical data using embeddings, or dense vector representations, allowing the model to differentiate between permafrost behavior in different latitudes. Most importantly, there are no limitations as to which numerical or categorical data can be fed into a DL model; as long as data is present and representable, DL models are able to extract the most important features without need for extensive domain knowledge.

This study introduces a deep learning pipeline for spatiotemporal modeling of Alaskan permafrost, aiming to forecast annual soil temperature profiles across four depth layers (0–0.07 m, 0.07–0.28 m, 0.28–1.00 m, and 1.00–2.89 m). Within this framework, five established deep learning architectures (TCN, Transformer, Conv1DLSTM, GRU, BiLSTM) were evaluated for predictive performance. A dataset was compiled from monthly aggregations of historical reanalysis data and static geological data corresponding to six latitudinal regions in Alaska. A pipeline was engineered to transform the time-series data into normalized sliding-window sequences and encode latitude with learned band embedding representations to improve spatial specificity. The models were then used to predict soil temperature profiles for future years across four representative concentration pathway (RCP) scenarios under the Coupled Model Intercomparison Project Phase 5 (CMIP5): RCP 2.6, RCP 4.5, RCP 6.0, and RCP 8.5. Finally, SHAP interpretability analysis was used to identify the most influential drivers of active layer behavior.
\section{Data and Methods}
\subsection{Study Region and Climate Context}
\begin{figure}[H]
  \centering
  \includegraphics[width=0.36\textwidth]{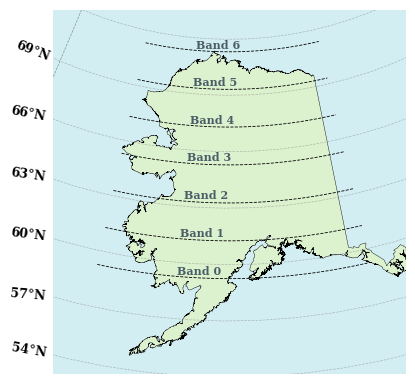}
  \caption{Band map of Alaska used in the analysis.}
  \label{fig:bandmap}
\end{figure}

The study focuses on the spatial extent of mainland Alaska from 59\textdegree N northwards, situated above the continuous permafrost belt, and is bounded by the northern frontier at 71.3\textdegree N. The complete region spans an area of 12.52 $\times 10^5$ square kilometers. The central Alaskan plateau constitutes the majority of this region, defined by hills and lowlands and a general altitude of 600--1500 meters \cite{Brooks1906}. Continuous permafrost in this zone is observed to be up to 650 meters thick \cite{Schuur2008}, with relatively stable thawing cycles. The area was partitioned into six latitude bands, each {2\textdegree} wide from north to south (\cref{tab:bands}, \cref{fig:bandmap}). 

\begin{table}[ht]
\footnotesize
\caption{Band indices and corresponding surface areas, upper latitude boundaries, and lower latitude boundaries.}\label{tab:bands}
\begin{center}
  \begin{tabular}{|c|c|c|c|} \hline
   \bf Band & \bf Surface Area (km$^2$) & \bf Upper Latitude Boundary & \bf Lower Latitude Boundary\\ \hline
    Band 0 & $2.43360 \times 10^5$ & 61.2755545\textdegree N & 59.2632802\textdegree N\\
    Band 1 & $2.64663 \times 10^5$ & 63.2878288\textdegree N & 61.2755545\textdegree N\\ 
    Band 2 & $2.38854 \times 10^5$ & 65.300103\textdegree N & 63.2878288\textdegree N\\ 
    Band 3 & $2.2582 \times 10^5$ & 67.3123774\textdegree N & 65.300103\textdegree N\\ 
    Band 4 & $2.12574 \times 10^5$ & 69.3246517\textdegree N & 67.3123774\textdegree N\\ 
    Band 5 & $6.7668 \times 10^4$ & 71.336926\textdegree N & 69.3246517\textdegree N\\ 
    \hline
  \end{tabular}
\end{center}
\end{table}

Active layer soil temperatures exhibit a declining trend with increasing latitude. ERA5-Land reanalysis data from 2001 to 2023 reports average monthly soil temperatures ranging from 3\textdegree C in the southern region to -2.3\textdegree C in the far northern region, at a depth of 0--0.07 meters \cite{MunozSabater2019}.

\subsection{Data Sources and Processing}
%\subsection{Data}
\subsubsection{Data Availability}
Active layer temperature hinges on a conglomerate of parameters \cite{vanHuissteden2020}. GIPL, for instance, relied on both static and time-dependent variables \cite{Jafarov2012}. In this study, compiled datasets include both forms of data. 

Deep learning entails large volumes of sequential data in order to learn long-term temporal dependencies. Large datasets of in situ observations, especially in Alaska, are often sparse and incomplete. Renalaysis datasets, on the contrary, provides continuous climate and environmental data with high spatial and temporal resolution. For this study, monthly soil temperature data over 23 years (2001--2023) corresponding to each latitude band was accessed from the ERA5-Land Reanalysis dataset under the European Centre for Medium-Range Weather Forecasts (ECMWF) through Google Earth Engine \cite{Gorelick2017}. ERA5-Land has a spatial resolution of 9 km (0.1\textdegree $\times$ 0.1\textdegree) and hourly temporal resolution \cite{MunozSabater2019}. For the purposes of this study, the data were aggregated to monthly intervals. Moreover, soil temperatures were fetched from shallow permafrost layers to zero in on the active layer domain. ERA5-Land offers soil temperature at four depth ranges: 0--0.07 m ($L_0$), 0.07--0.28 m ($L_1$), 0.28--1.00 m ($L_2$), and 1.00--2.89 m ($L_3$). The layer depths can be arithmetically represented as a vector $\vec{L} = \langle L_0, L_1, L_2, L_3\rangle$, and their corresponding monthly temperatures by latitude band as $T:\vec{L} \xrightarrow{} \vec{T}_{b,t} = \langle T_{b,t,0} ,T_{b,t,1} ,T_{b,t,2}, T_{b,t,3} \rangle$, where $T_{b,t,i}$ is the calculated soil temperature (in \textdegree C)  for band $b$, time step $t$, and layer depth range $i$. The static data, on the other hand, can be tabulated as individual values per latitude band. The Geologic Map of Alaska from the United States Geological Survey (USGS) provides surveyed lithological data by geospatial coordinates in a 1:250,000 scale, and was used to obtain lithological aggregations for the state of Alaska \cite{Wilson2015}. Additional static data were retrieved from OpenLandMap, which has a 250--meter resolution \citep{Hengl2018_BulkDensity, Hengl2018_SOC}. 

Soil temperature trajectories required input features derived from matching simulated future data. CMIP5 is a coordinated climate modeling program under the Program for Climate Model Diagnosis and Intercomparison (PCMDI) which offers simulations under multiple scenarios and Earth System Models (ESMs) \citep{PCMDI, Taylor2012}. This study retrieved data on RCP 2.6, RCP 4.5, RCP 6.0, and RCP 8.5 from the IPSL-CM5A-MR model developed under the Institute of Pierre-Simon Laplace. IPSL-CM5A-MR (Version 5) is a comprehensive ESM with a 2.5\textdegree $\times$ {1.25\textdegree} horizontal atmospheric resolution and a vertical atmospheric resolution of 39 levels \cite{Dufresne2013}.

\subsubsection{Data Preparation}
Climate, environmental, temperature, and geophysical features were evaluated before extracting them from datasets, particularly for dynamic data. Availability under both the ERA5-Land reanalysis dataset and IPSL-CM5A-MR’s simulated results determined which features could be extracted. Summer air temperature levels ($T_{AIR}$) and precipitation amount were observed to directly increase soil heat content in the active layer \cite{Schuur2008}. Precipitation in the cryosphere can be categorized into total precipitation ($P_{TOTAL}$) and snowfall amount ($P_{SNOW}$). Conceptual models suggest that net radiation balance has a direct effect on soil and surface temperatures \citep{Nguyen2020, vanHuissteden2020}. ALT has been reported to thicken from longer thawing cycles due to surface temperature ($T_{SURFACE}$) and downwelling longwave radiation ($R_{THERMAL}$) \cite{Liu2024} . Moreover, latitude affects the distribution of insolation, partially influenced by downwelling shortwave radiation ($R_{SOLAR}$) onto the planet’s surface \cite{vanHuissteden2020}. The Budyko--Seller’s model asserts that absorbed insolation is a parameter in determining surface temperature \citep{Budyko1969, Nguyen2020}. Energy fluxes are also relevant in the warming of surface and soil temperatures \cite{vanHuissteden2020}. In particular, latent heat flux ($Q_{LATENT}$) and sensible heat flux ($Q_{SENSIBLE}$) (as well as ground heat flux , which is unavailable in ERA5-Land) are crucial components of net radiation balance. All of the energy balance variables are represented in $J\cdot m^{-2}$. Lastly, the zero-curtain effect during the transition between freezing and thawing is contingent on volumetric water content (accounting for both liquid groundwater and ice) in the soil \cite{Muniz2025}. Energy flow simulations have validated the positive correlational relationship between volumetric water content and longevity of thawing period \cite{Muniz2025}. This study represents layer-wise volumetric water content as the vector result of the linear transformation $T:\vec{L} \xrightarrow{} \vec{W}_{b, t} = \langle W_{b,t,0}, W_{b,t,1}, W_{b,t,2}, W_{b,t,3}\rangle$ where $W_{b,t,i}$ is the volume fraction of water in meters for band $b$, time step $t$, and layer depth range $i$.

All datasets were compiled and processed in Python using Jupyter notebooks executed in Google Colaboratory \cite{googlecolab}. Daily reanalysis data for 23 years (2001-2023) were extracted per latitude band $b$ for the features $T_{AIR}$, $T_{SURFACE}$, $P_{TOTAL}$, $P_{SNOW}$, $R_{THERMAL}$, $R_{SOLAR}$, $Q_{LATENT}$, $Q_{SENSIBLE}$, $\vec{W}_{b,t}$ from ERA5-Land, compiled into a dataset, and aggregated into monthly intervals \cite{Gorelick2017}. In addition, the corresponding year, sine value of the month index, and cosine value of the month index, were compiled in accordance with the monthly feature values. Time-invariant data were summarized per band. Lithologic classes (water, unconsolidated surficial deposits, glacier, volcanic, melange, plutonic, sedimentary, metamorphic, and unknown) were aggregated by percentage for each latitude band $b$ from USGS’s Geologic Map of Alaska \cite{Wilson2015}. Bulk density data ($B_{b,L}$) in $kg\cdot m^{-3}$, and organic carbon content data ($C_{b,L}$) in $g\cdot kg^{-1}$ for each band $b$ and each soil layer $L \in \vec{L}$ were also aggregated \citep{Hengl2018_BulkDensity, Hengl2018_SOC}. The dataset will be represented as a matrix $X_{hist} \in  \mathbb{R}^{n \times F}$ , where $n$ is the total number of monthly historical time steps and $F$ is the total number of features. Corresponding target values of $T_{b,t}$ were compiled as the dataset $Y \in \mathbb{R}^{n \times 4}$. 

Scenario data had to be compiled in a similar manner from IPSL-CM5A-MR \cite{xarray}. Available parameters in the CMIP5 archive were manually mapped as proxies of ERA5-Land in order to maintain consistency \cite{PCMDI_CMIP5_Standard_Output_2009}. Datasets containing the previously mentioned features for the years 2024–2030 were organized for each scenario, then merged with band-wise static data to ensure parallelism with historical data. The resulting datasets can be depicted with the following symbols: $\hat{X}_{2.6}$, $\hat{X}_{4.5}$, $\hat{X}_{6.0}$, and $\hat{X}_{8.5} \in \mathbb{R}^{m \times F}$, where $m$ is the number of monthly scenario-specific time steps. IPSL-CM5A-MR also offers simulated historical data for the previous features, which were also compiled for the years 2001–2023, excluding all static data. This additional historical dataset will be depicted as $\hat{X}_{hist} \in \mathbb{R}^{n \times F_d}$, where $F_d$ represents the number of dynamic features. 

Appropriate unit conversions were done for the values in $\hat{X}_{2.6}$, $\hat{X}_{4.5}$, $\hat{X}_{6.0}$, $\hat{X}_{8.5}$, and $\hat{X}_{hist}$ to match with that of $X_{hist}$. To estimate scenario-specific volumetric water content by layer, a Dirichlet-based algorithm was carried out, derived from a combination of soil moisture data from IPSL-CM5A-MR and corresponding layer-wise volumetric soil water content variables from ERA5-Land. 

Finally, a bias-scaling procedure was executed to eliminate systemic biases in scenario-specific data. Noel et al. \cite{NOEL2021106900} performed a downscaling procedure called Quantile Mapping (QM) which compared the distributional differences between historical CMIP5 data and ERA5-Land Reanalysis data to learn bias correction mapping. This mapping was applied to adjust CMIP5 future projections. In this study,  $\hat{X}_{hist}$ and $X_{hist}$ were used for producing mappings, and the adjustments were carried out onto $\hat{X}_{2.6}$, $\hat{X}_{4.5}$, $\hat{X}_{6.0}$, and $\hat{X}_{8.5}$. The final, clean RCP datasets are now $X_{2.6}$, $X_{4.5}$, $X_{6.0}$, $X_{8.5}$.

\paragraph{Scenario Signal}
This study implemented a scenario signal feature, denoted as $z$, to emphasize climate forcing intensity between RCP scenarios. The deep learning models cannot naively be trained to learn differences in distributional alignment or intensity of RCP scenarios since scenario-specific target data on soil temperatures are not provided by CMIP5. Furthermore, the models were originally purposed to predict scenario-specific soil temperatures; thus, the lack of scenario training may pose a serious hindrance. For this reason, an additional feature $z$ was derived as a normalized anomaly of the mean of $R_{THERMAL}$. These changes were applied to all datasets: $X_{hist}$, $X_{2.6}$, $X_{4.5}$, $X_{6.0}$, and $X_{8.5}$. The mean and standard deviation of historical $R_{THERMAL}$ was computed as $\mu_{THERMAL}$ and $\sigma_{THERMAL}$ respectively. In addition, the mean of $R_{THERMAL}$ of a window sequence $j$ was computed as $\bar{x}_{THERMAL}^j$. Sliding window sequences will be detailed in \cref{sliding-windows}. The final $z_j$ feature of $j$ was calculated by \cref{eq:scensignal_1} and \cref{eq:scensignal_2}.

\begin{equation}\label{eq:scensignal_1}
\tilde{z}_j = \frac{\bar{x}_{THERMAL}^j - \mu_{THERMAL}}{\sigma_{THERMAL}}
\end{equation}

\begin{equation}\label{eq:scensignal_2}
z_j = |\tilde{z}_j| * \tilde{z}_j
\end{equation}

\subsubsection{Temporal Windowing for Sequential Modeling}
\begin{figure}[htbp]
  \centering
  \includegraphics[width=0.9\textwidth]{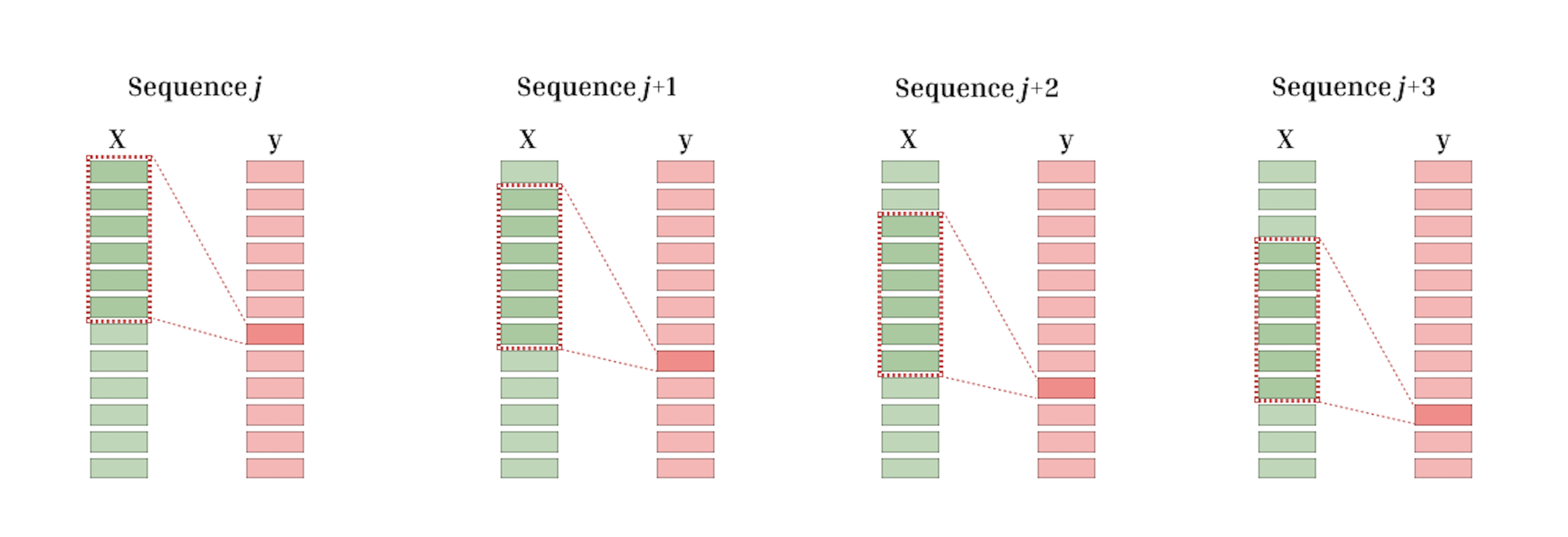}
  \caption{Diagram of sliding windows for sequence $j$ and subsequent sequences.}
  \label{fig:slidingwindows}
\end{figure}

%\subsubsection{Sliding Window Sequences}
\label{sliding-windows} A key architectural advantage of deep learning is its ability to learn temporal dependencies by sequentially processing multi-dimensional inputs. In this study, the soil temperature vector $\vec{T}_{b, m, y}$ will be predicted with input data from the previous 24 months, creating a window of seasonal context for the models to analyze, and enabling it to learn dynamic climate patterns such as lag effects and periodicity. A sliding-window approach was taken to transform $X$ \footnote{In this paper, references to $X$ should be understood to apply to $X_{hist}$, $X_{2.6}$, $X_{4.5}$, $X_{6.0}$, and $X_{8.5}$} and $Y$ into time step sequences for a supervised learning format with a fixed window size $w = 24$ (\cref{fig:slidingwindows}). For each band $b$, data was sorted chronologically. Each time step is labeled as $t \in [0,  n–w]$, where $t$ is the unique index for a time step at month $m$ and year $y$, and each sequence  $j_b \in \{0, …, N–1\}$ where $N$ is the total number of generated sequences for a band $b$, the pipeline observes the three subsequences \cref{eq:subsequence_x}, \cref{eq:subsequence_b}, and \cref{eq:subsequence_y}.

\begin{equation}\label{eq:subsequence_x}
\mathcal{X}_{j,b} = [x_t, x_{t+1},...,x_{t+w-1}] \in \mathbb{R}^{w \times F} 
\end{equation}
\begin{equation}\label{eq:subsequence_b}
\mathcal{B}_{j,b} = [b, b, ...,b] \in \mathbb{R}^w
\end{equation}
\begin{equation}\label{eq:subsequence_y}
\mathcal{Y}_{j,b} = T_{b,t+w} \in \mathbb{R}^4
\end{equation}

$\mathcal{X}_{j,b}$ is the $j$-th input window consisting of augmented feature vectors $x_i \in X$ for time step $i \in \{t, …, t+w–1\}$, in window size $w$. Each $\mathcal{X}_{j,b}$ has a corresponding $\mathcal{Y}_{j,b}$ representing the soil temperature vector for the next consecutive month $t+w$ after the 24-month sequence. 

The historical dataset $X_{hist}$ was split into a training dataset $X_{train}$ and a testing dataset $X_{test}$. The corresponding target data $Y$ was split likewise into $Y_{train}$ and $Y_{test}$. Since deep learning models learn sequentially, train-test split could not be random; hence, the first 18 years were assigned to the training dataset, and the last 5 years were assigned to the testing dataset. Sequences \cref{eq:sequence_h} and \cref{eq:sequence_s} were then generated for training data, testing data, and for each scenario data.

\begin{equation}\label{eq:sequence_h}
\mathcal{S}_{h} = T:(X_{h}, Y_{h}) \xrightarrow{} \bigcup_{b=0}^{B-1} \{{(\mathcal{X}_{j,b}, \mathcal{B}_{j,b}, \mathcal{Y}_{j,b})\}^{N_{h}}_{j=1}}_{h} ,
h \in \{train, test\}
\end{equation}

\begin{equation}\label{eq:sequence_s}
\mathcal{S}_{s} = T:X_{s} \xrightarrow{} \bigcup_{b=0}^{B-1} \{{(\mathcal{X}_{j,b}, \mathcal{B}_{j,b}, \mathcal{Y}_{j,b})\}^{N_{s}}_{j=1}}_{s}, 
s \in \{2.6, 4.5, 6.0, 8.5\}
\end{equation}

\paragraph{Additional Preprocessing}
All features in $\mathcal{S}_{h}$ and $\mathcal{S}_{s}$ were z-score normalized using the mean and standard deviation of training set \cite{Pedregosa2011}.

\subsection{Data Pipeline}

\subsubsection{Deep Learning Models}

\paragraph{Convolutional Neural Networks (CNNs)}

Convolutional neural networks apply convolutional operations to input data using a superimposed sliding kernel matrix \cite{Lecun1989}. They notably excel in pattern detection, making it a commonly used tool for image classification. In the context of climate modeling, convolutional layers can be used to detect seasonal patterns and spatiotemporal locality. This study uses two types of convolutional neural networks: a Temporal Convolutional Network (TCN) and a Convolutional 1D Long-Short Term Memory (Conv1DLSTM). The latter will be detailed in the next section. Temporal Convolutional Networks work by integrating causal and dilated convolutions, which prevents leakage of future data during the learning process, and expands the receptive field of the model \cite{bai2018}. This maintains model integrity and enables it to learn both short-term and long-term temporal patterns. Additionally, soil temperature in particular is prone to local shifts from short-to-medium range temporal contexts; a TCN, in this case, would excel in capturing spatiotemporal local patterns. Permafrost dynamics also hold long-term memory from built-up heat, and a TCN would be able to simulate this behavior through dilated convolution.

\paragraph{Recurrent Neural Networks (RNNs)}

Recurrent neural networks were originally designed to process sequential data, and are distinguished by holding memory of previous sequences and capturing a lag effect \cite{Elman1990}. The lag effect allows RNNs to learn delayed influences of past sequences on present temperatures. Two forms of RNNs have emerged: the Long-Short Term Memory (LSTM) and the Gated Recurrent Unit (GRU) \citep{Chung2014, hochreiter1997}. This study uses a GRU, the simpler of the two, for its ability to filter out important events from memory, while forgetting irrelevant past events, essentially preventing it from being noise-sensitive while also capturing big-picture temperature trends \cite{Chung2014}. LSTMs, on the other hand, are more complex to a degree, but have have stronger memory control compared to a GRU \cite{hochreiter1997}. Two LSTM architectures are trained in this study: a Bidirectional LSTM (BiLSTM), and a Convolutional 1D LSTM (Conv1DLSTM, as previously mentioned). Bidirectional LSTMs run two LSTMs: one which learns sequentially from past to future, and another which learns sequentially from future to past. This form of LSTM is particularly powerful in understanding reverse-lag effects, or how the future may be used to map the past \cite{schuster1997}. Lastly, a Conv1DLSTM is a hybrid between CNNs and RNNs which jointly extracts local short-term feature patterns through its one-dimensional convolutional layer while also learning long-range dependencies from its LSTM layer \cite{ordonez2016}. 

\paragraph{Transformers}

Transformers use self-attention mechanisms to understand relationships between time steps and features in terms of their relevance to one another, and positional encoding to learn cyclic temporal patterns \cite{vaswani2023attentionneed}. Although they were originally designed for natural language processing, transformers are also applicable when understanding time series. Transformers would be notably powerful in climate modeling due to their ability to process timesteps in parallel. In a sequential climate model, self-attention allows transformers to focus on key months to learn seasonal recurrences; iteratively adjusted attention weights also allow the transformer to dynamically learn feature importance. 

\subsubsection{Model Initialization and Setup}
\begin{figure}[htbp]
  \centering
  \includegraphics[width=0.9\textwidth]{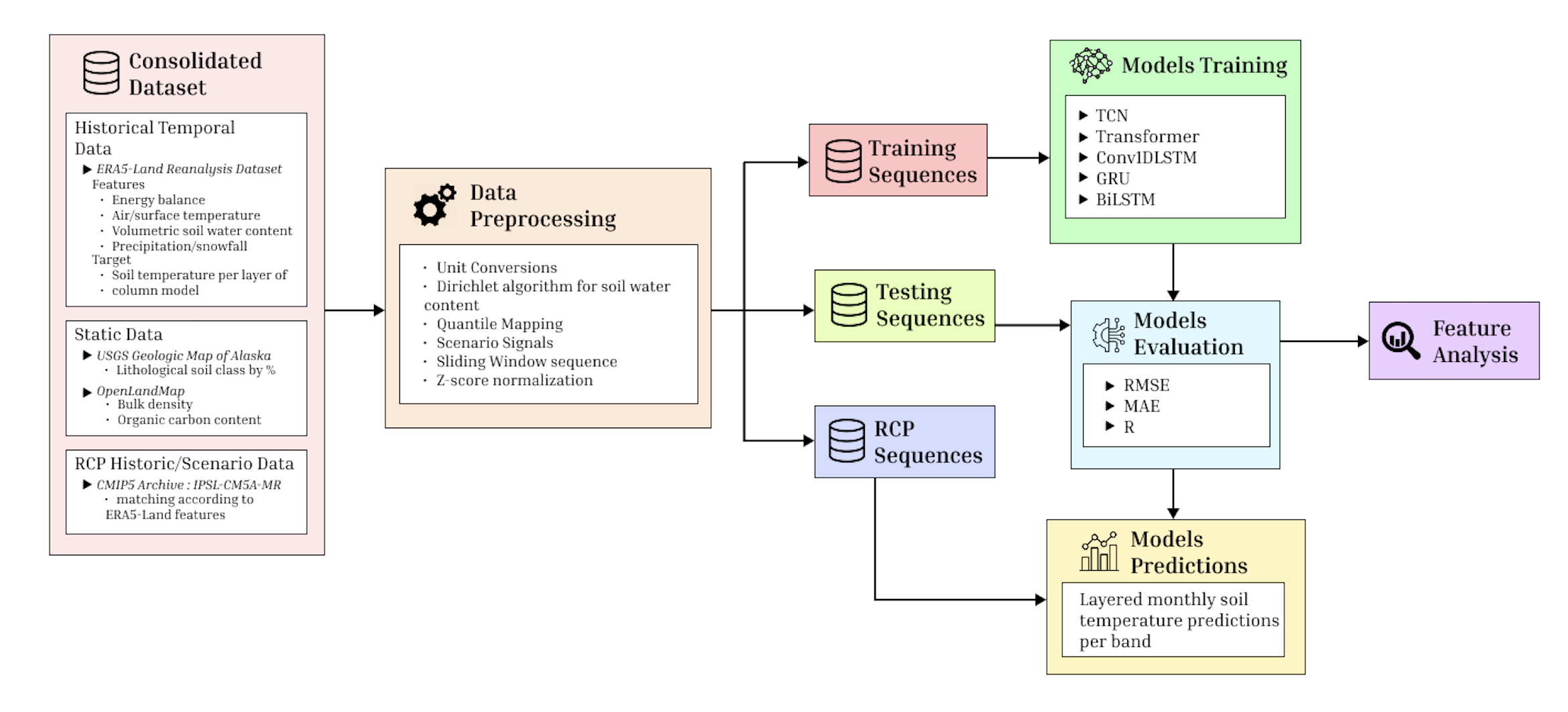}
  \caption{Proposed end-to-end deep-learning workflow of model training, testing, and final soil temperature predictions.}
  \label{fig:pipeline}
\end{figure}

Model training, evaluation, and subsequent feature analysis (\cref{sec:feature_analysis}) were executed on  Jupyter notebooks in Google Colaboratory as well \cite{googlecolab}. Each model was trained for 50 epochs on $\mathcal{S}_{train}$ (2001–2018) and evaluated on $\mathcal{S}_{test}$ using RMSE, MAE, and $\mathrm{R}^2$ scores. All models include learnable band embeddings $e_{b,j} \in \mathbb{R}^4$, allowing the networks to capture spatial dependencies across different latitude bands. In this study, embeddings are used to map the categorical band index sequences $\mathcal{B}_{j,b}$ into dense vector representations, allowing the model to internalize parallels and distinctions between bands rather than treating them as fixed identifiers. Trained models were then applied to RCP scenario data ($\mathcal{S}_{2.6}$, $\mathcal{S}_{4.5}$, $\mathcal{S}_{6.0}$, and $\mathcal{S}_{8.5}$) to generate soil temperature projections for 2024-2030. The full pipeline is visualized in \cref{fig:pipeline}.

\subsection{Feature Analysis}

Feature analysis evaluates the degree of influence each feature has on a model’s prediction, and provides insight on the main drivers of active layer soil temperatures \cite{WANG2023165709}. This study employs SHapley Additive exPlanations (SHAP) to quantify the contribution of each feature to soil temperature predictions \cite{Lundberg}. For each model, this study sampled 100 sequences from $\tilde{{\mathcal{S}}}_{test}$ to produce baseline model predictions, and 10 additional sequences to determine feature contribution relative to the baseline predictions, outputting a SHAP score for each feature, which measures its relative magnitude of contribution. This algorithm was repeated over every band and layer, and the per-feature SHAP scores were aggregated to produce one SHAP score per feature for a given model.

\section{Results \& Discussion}

\subsection{Performance Analysis}
The performance of the deep learning models was evaluated for the years 2019-2023. The predictions for those years were compared with their respective ground truth using RMSE, MAE, and $\mathrm{R}^2$ scores \cite{Pedregosa2011}. \cref{tab:scores} details the individual scores for each model. The model ensemble was observed to have a close range of scores. The GRU had the strongest overall performance out of the five models. The RMSE scores of all models hover between 1.0 and 1.3, which is moderately low in the domain of soil temperatures, that range between -5\textdegree C to 10\textdegree C. The GRU holds the lowest RMSE score, implying there were fewer overall significant errors observed for this model. On the other hand, the BiLSTM predicted the greatest number of significant errors, with the largest RMSE score of 1.234. The MAE score results tell a similar story: the GRU had the lowest MAE score of 0.747. The underperformer, however, is the Transformer network, which had a score of 0.912. Here, the GRU also displayed the highest $\mathrm{R}^2$ score of 0.943, suggesting the best proportion of variance. Meanwhile, the BiLSTM showed the weakest $\mathrm{R}^2$ score. Nevertheless, the narrow range of RMSE, MAE, and $\mathrm{R}^2$ scores for all models suggests that all models are highly explanatory in the realm of temperature modeling. 

The next best overall performer besides the GRU was the Conv1DLSTM model. The Conv1DLSTM had the third highest $\mathrm{R}^2$ score, and the second lowest RMSE/MAE scores. Both the GRU and the Conv1DLSTM have architectural components which play a role in filtering out noise (in this case, it would be the update gate for the GRU, and the forget gate or convolutional layer for the Conv1DLSTM), explaining lower RMSE and MAE scores. The GRU may have outperformed in particular due to its simpler architecture and fewer weights, minimizing the chance of overfitting during backpropagation. Meanwhile, the BiLSTM may have underperformed due to its bidirectional nature, which may be a disadvantage in causal time-series forecasting due to it not relying solely on past information. 

\begin{table}[htbp]
\footnotesize
\caption{Model performances scores using RMSE, MAE, and $\mathrm{R}^2$.}\label{tab:scores}
\begin{center}
  \begin{tabular}{|c|c|c|c|} \hline
   \bf Model & \bf RMSE & \bf MAE & \bf $\mathrm{R}^2$\\ \hline
    TCN & 1.118 & 0.872 & 0.937 \\
    Transformer & 1.158 & 0.912 & 0.939 \\ 
    Conv1DLSTM & 1.092 & 0.833 & 0.938 \\ 
    GRU & 1.027 & 0.747 & 0.943 \\ 
    BiLISTM & 1.234 & 0.907 & 0.919 \\ 
    \hline
  \end{tabular}
\end{center}
\end{table}

\begin{figure}[htbp]
  \centering
  \includegraphics[width=0.45\textwidth]{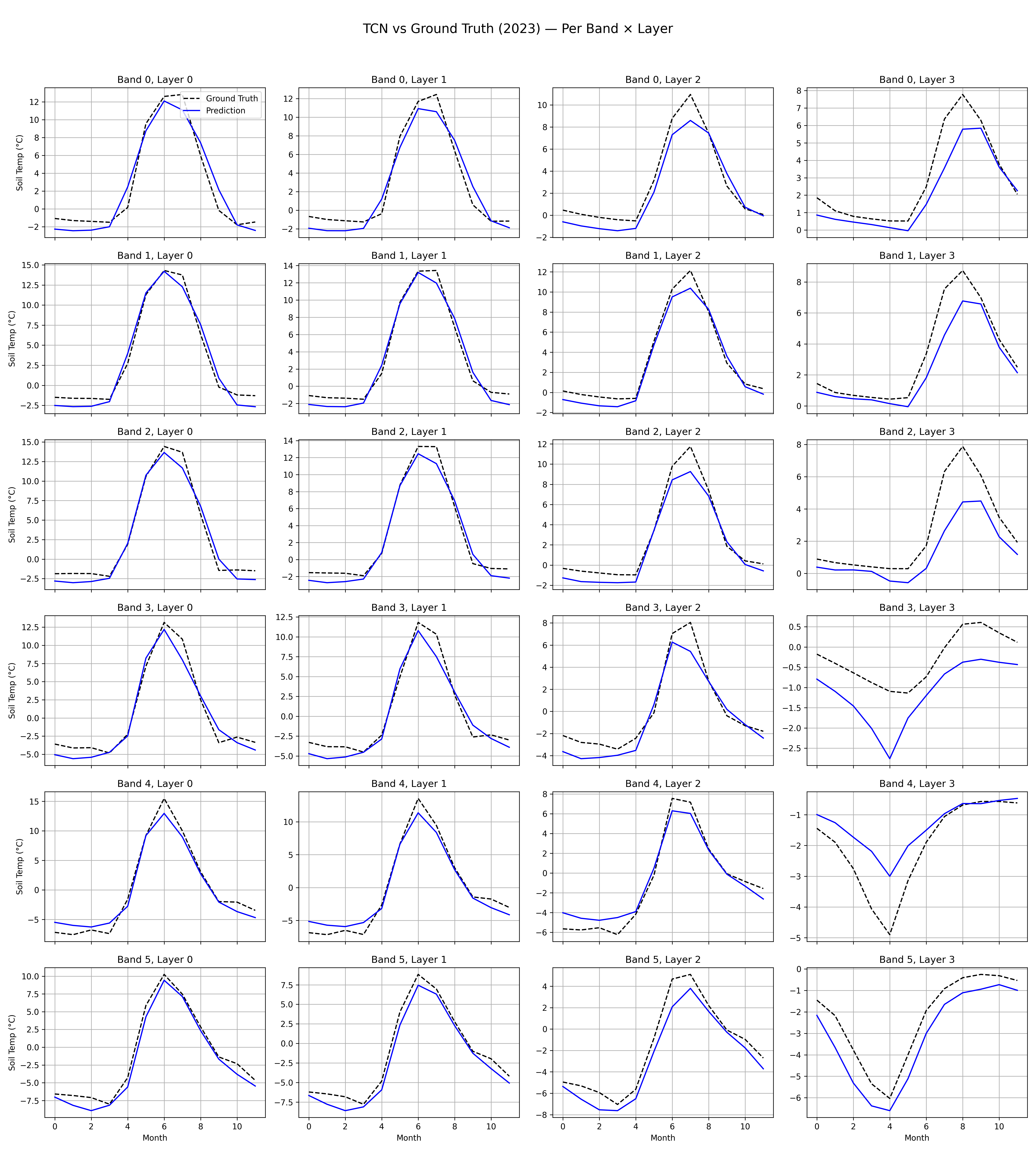}
  \includegraphics[width=0.45\textwidth]{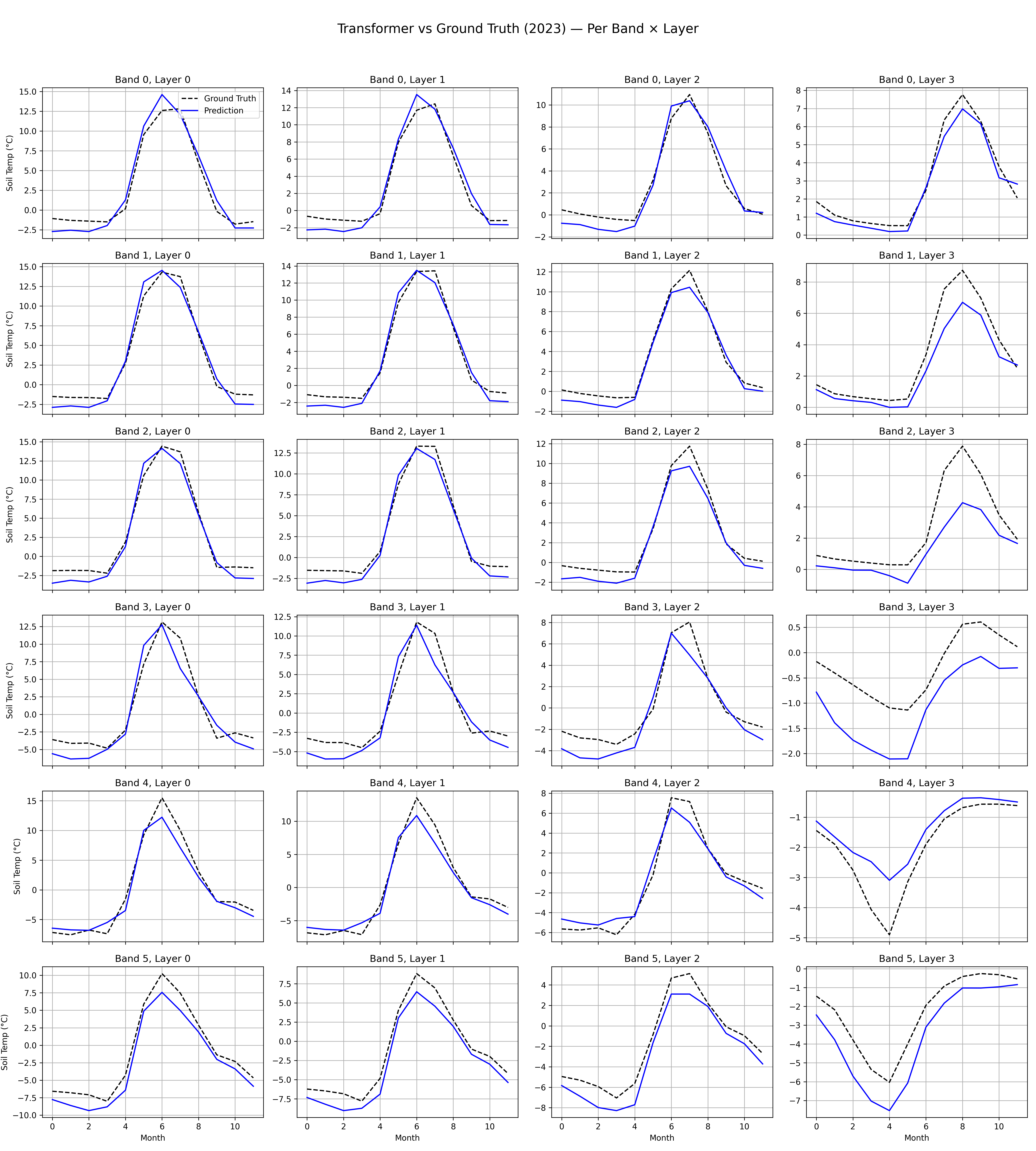}
  \includegraphics[width=0.45\textwidth]{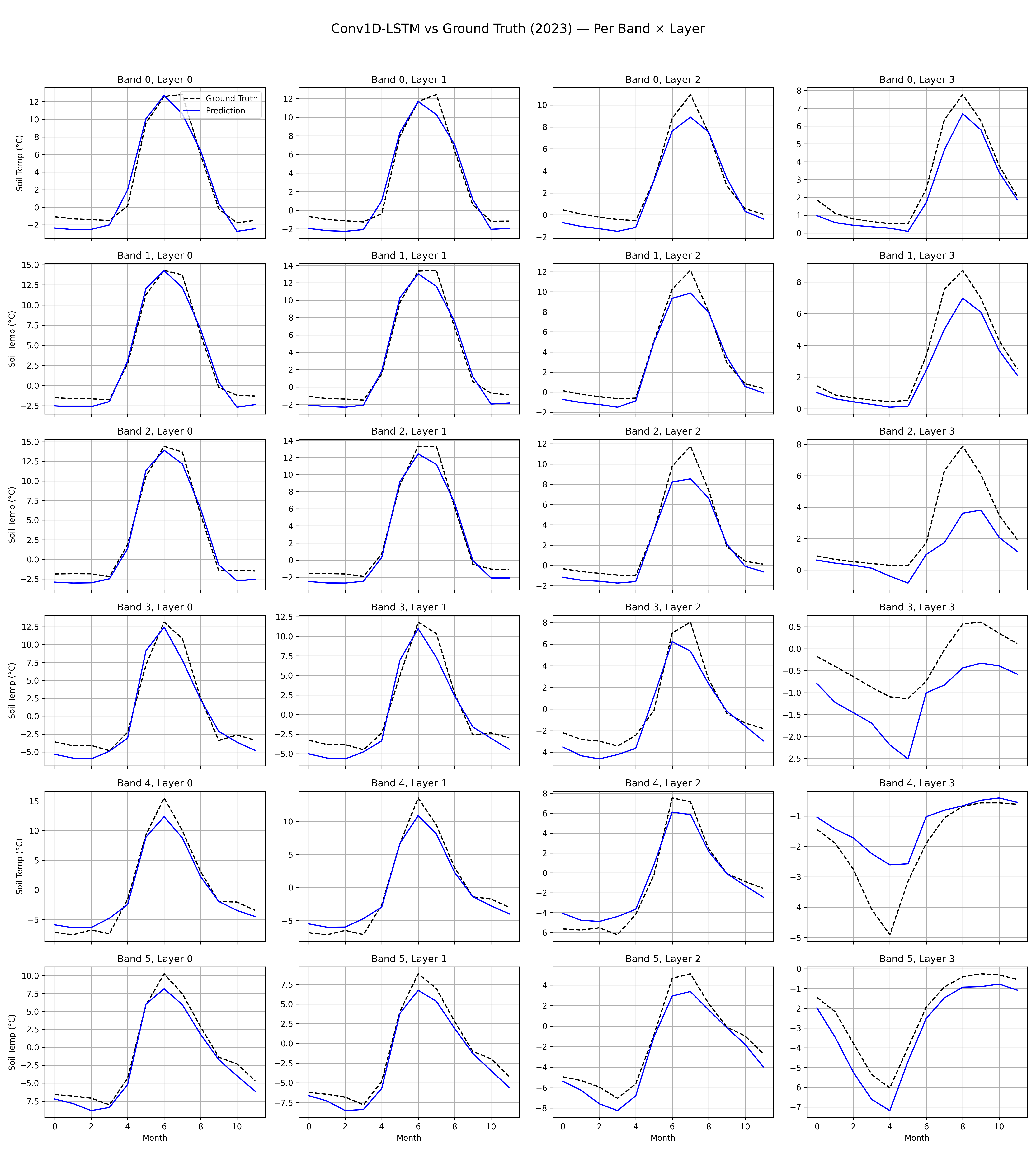}
  \includegraphics[width=0.45\textwidth]{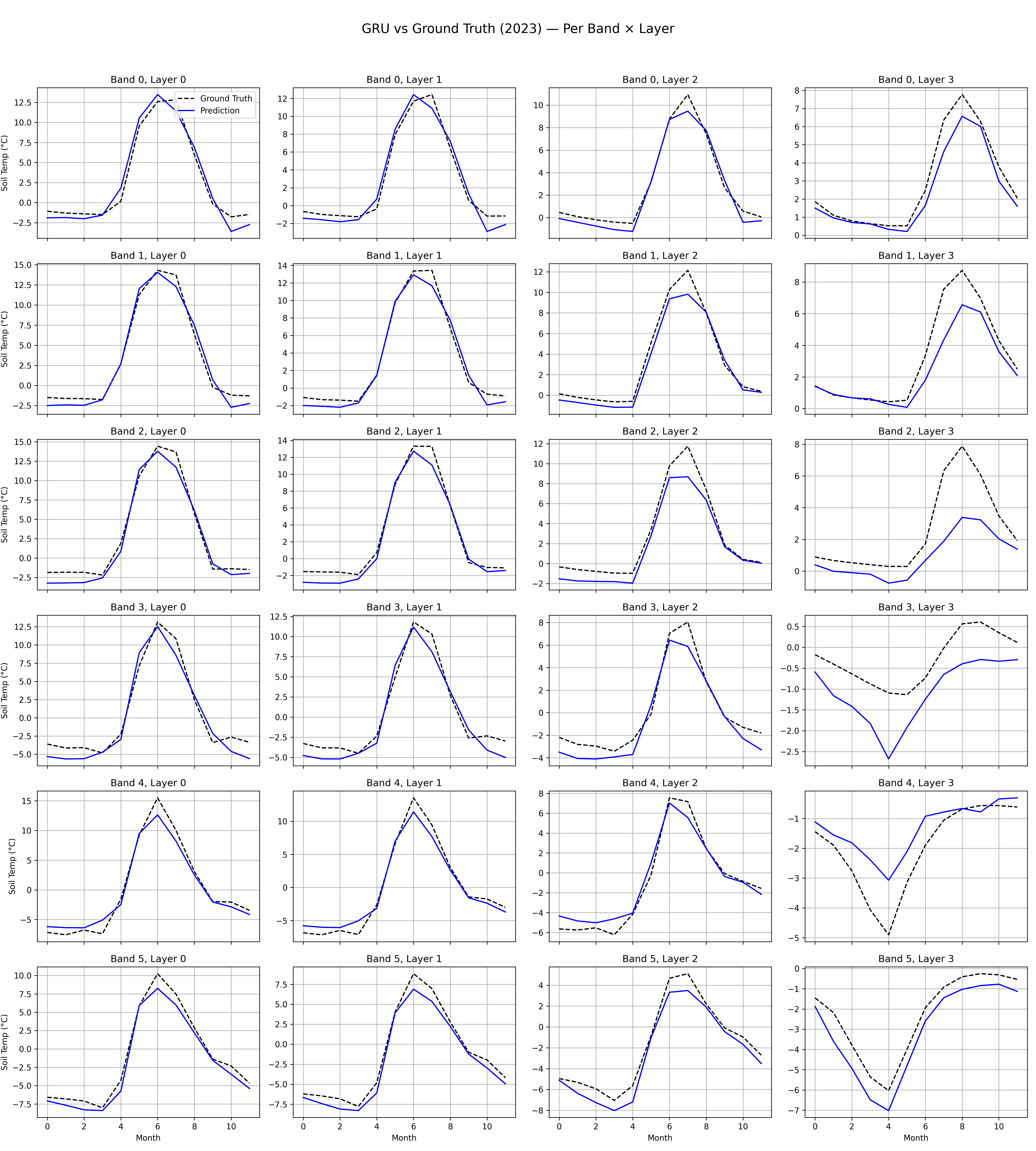}
  \includegraphics[width=0.45\textwidth]{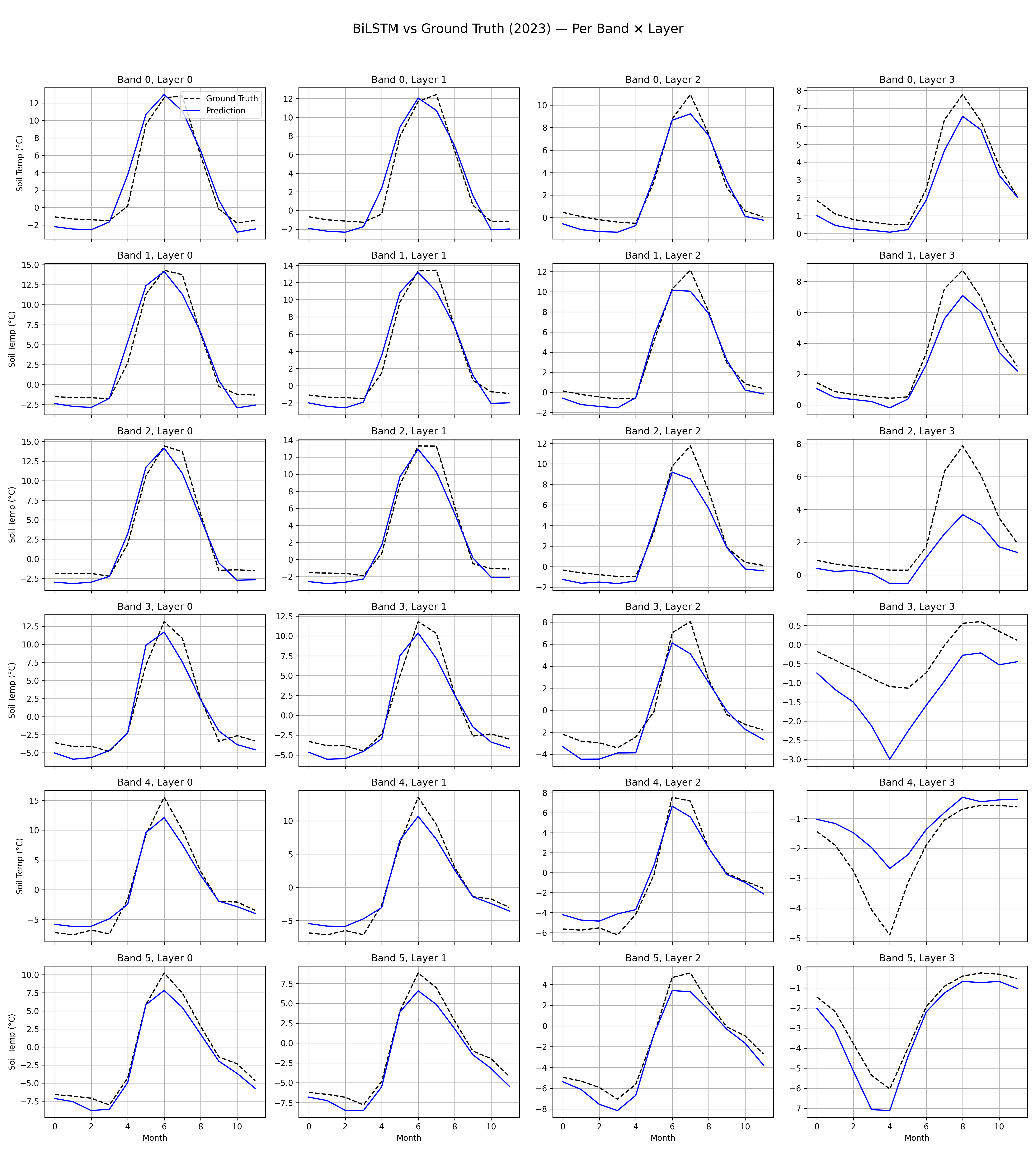}
  \caption{Comparison of model predictions (blue) with ground truth (dashed) for the year of 2023, stratified by latitude band (rows) and soil layer (columns).}
  \label{fig:evaluation}
\end{figure}

In order to better analyze the performance of each model, predictions were validated using ground truth for the year 2023. Note that the models were not trained for the year 2023, so the inputs for this year would essentially be foreign information, ensuring fair evaluation on their performances. \cref{fig:evaluation} displays the plots for observed versus predicted soil temperature for all models, bands, and layers.

Across the 120 prediction plots (24 per model), strikingly consistent behavior was observed. Rather than drawing conclusions from each individual graph, the analysis below highlights overarching trends and points of divergence to represent the general performance of the model ensemble. 

From a general perspective, all five models exhibited a profound understanding of sinusoidal seasonal variations, aligning very similarly with the summer crests and winter troughs of ground truth temperatures. Very rarely were predictions shifted horizontally from the ground truth, implying that all members of the model ensemble learned short-term consistencies in seasonal temperature patterns. Moreover, the visual correspondence between predicted temperatures and actual temperatures for all bands and layer depths suggests that all five models have learned spatial and vertical differences. For instance, all models were able to capture the gradual decrease of the lower bound during winter seasons as the band number increases, which mirrors the natural northward trend of progressively colder temperatures; this further suggests that band embeddings were effective in distinguishing latitudinal variations. Another common area of strength for nearly all model predictions, particularly for the first three layer depth ranges, were the near-overlaps spring and fall seasons, when the temperatures were either increasing or decreasing. Such performance suggests all models were able to learn short-term gradients and long-term periodicity of soil temperature from the given environmental and thermal data. 

However, a unanimous weakness for all models was at layer depth 3 (1.00--2.89 m), particularly for higher latitudes. Although the general seasonal trend was captured, the absolute difference between the extrema was significant, explaining why RMSE scores for all five models were greater than 1. This can be explained by layer 3 of bands 2, 3, 4, and 5 exhibiting different seasonal patterns than the other layers and bands, with freezing soil temperatures stretching through spring. The delayed thawing pattern in the layer depth 3 may have been due to thermal inertia or snow insulation, which were not included as features for this study, explaining why the models were not able to learn a different seasonal trend for the lowest layer \cite{Schuur2008}. 

\subsection{Scenario Forecast Results}

\begin{figure}[htbp]
  \centering
  \includegraphics[width=0.45\textwidth, angle=-90]{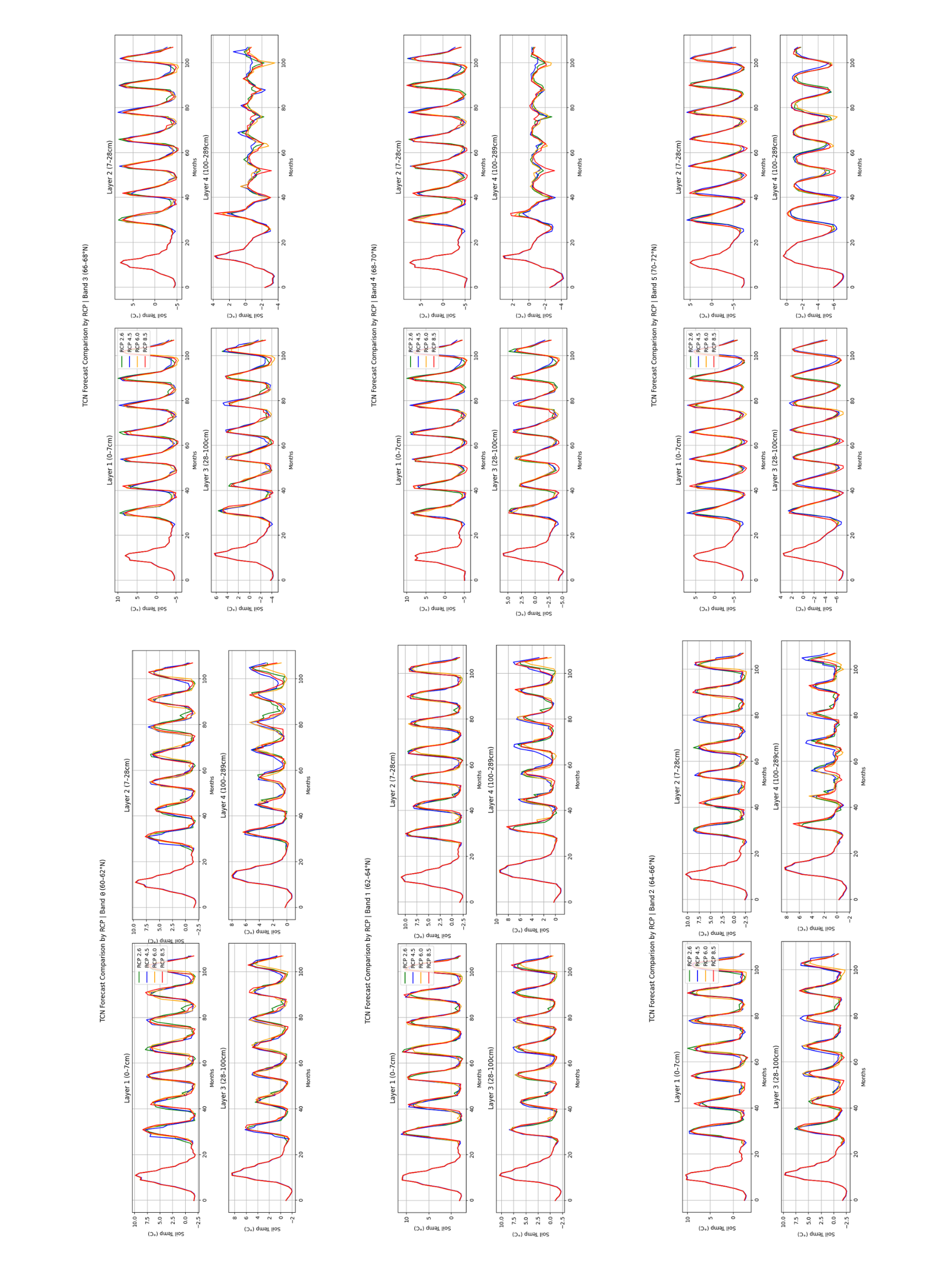}
  \label{fig:predictions_tcn}
  \caption{Comparison between TCN predictions per band-layer pair for each RCP scenario. RCP 2.6 is in green, RCP 4.5 is in blue, RCP 6.0 is in yellow, and RCP 8.5 is in red.}
\end{figure}

\begin{figure}[htbp]
  \centering
  \includegraphics[width=0.45\textwidth, angle=-90]{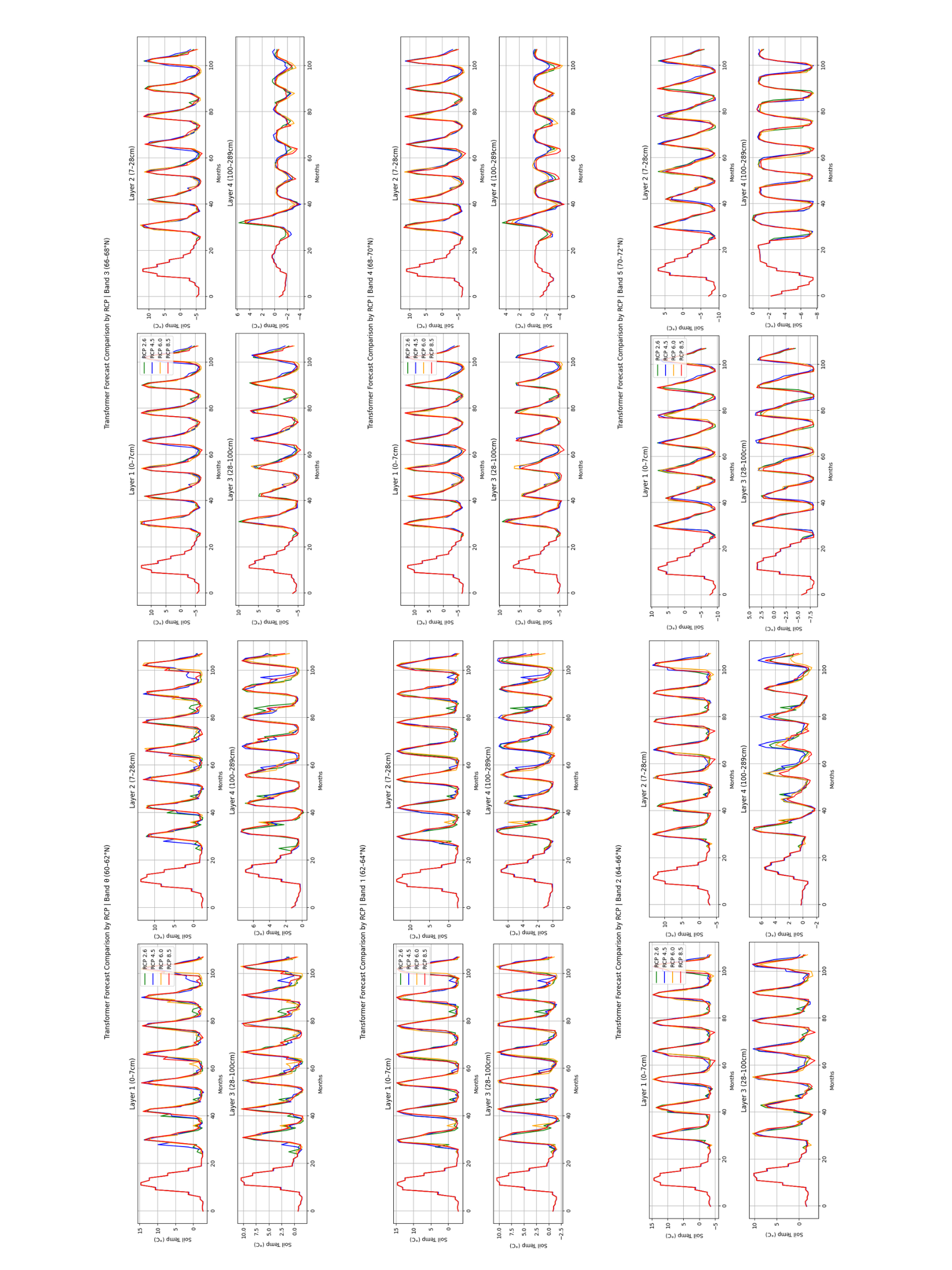}
  \caption{Comparison between Transformer predictions per band-layer pair for each RCP scenario. RCP 2.6 is in green, RCP 4.5 is in blue, RCP 6.0 is in yellow, and RCP 8.5 is in red.}
  \label{fig:predictions_trans}
\end{figure}

\begin{figure}[htbp]
  \centering
  \includegraphics[width=0.45\textwidth, angle=-90]{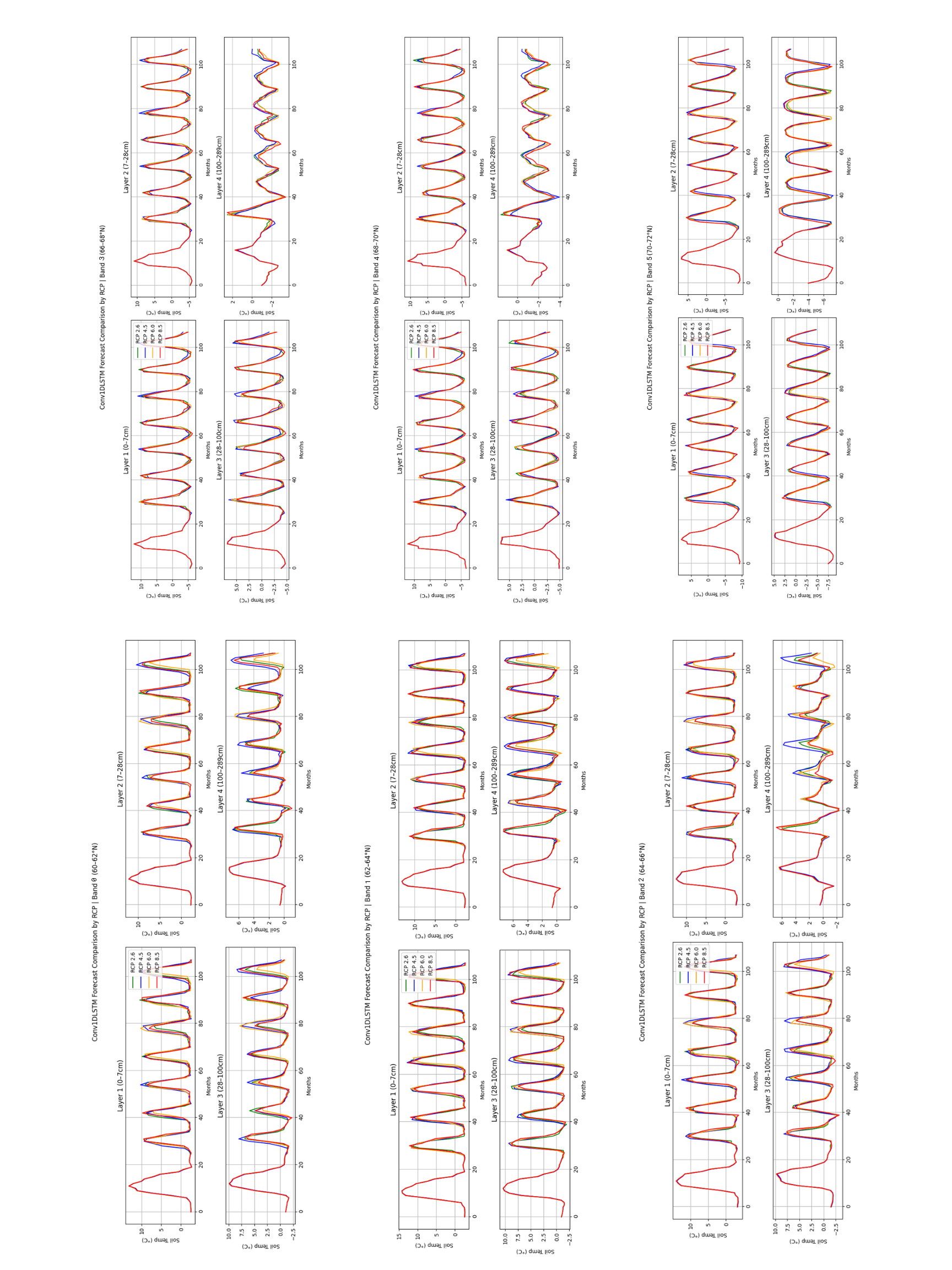}
  \caption{Comparison between Conv1DLSTM predictions per band-layer pair for each RCP scenario. RCP 2.6 is in green, RCP 4.5 is in blue, RCP 6.0 is in yellow, and RCP 8.5 is in red.}
  \label{fig:predictions_conv1dlstm}
\end{figure}

\begin{figure}[htbp]
  \centering
  \includegraphics[width=0.45\textwidth, angle=-90]{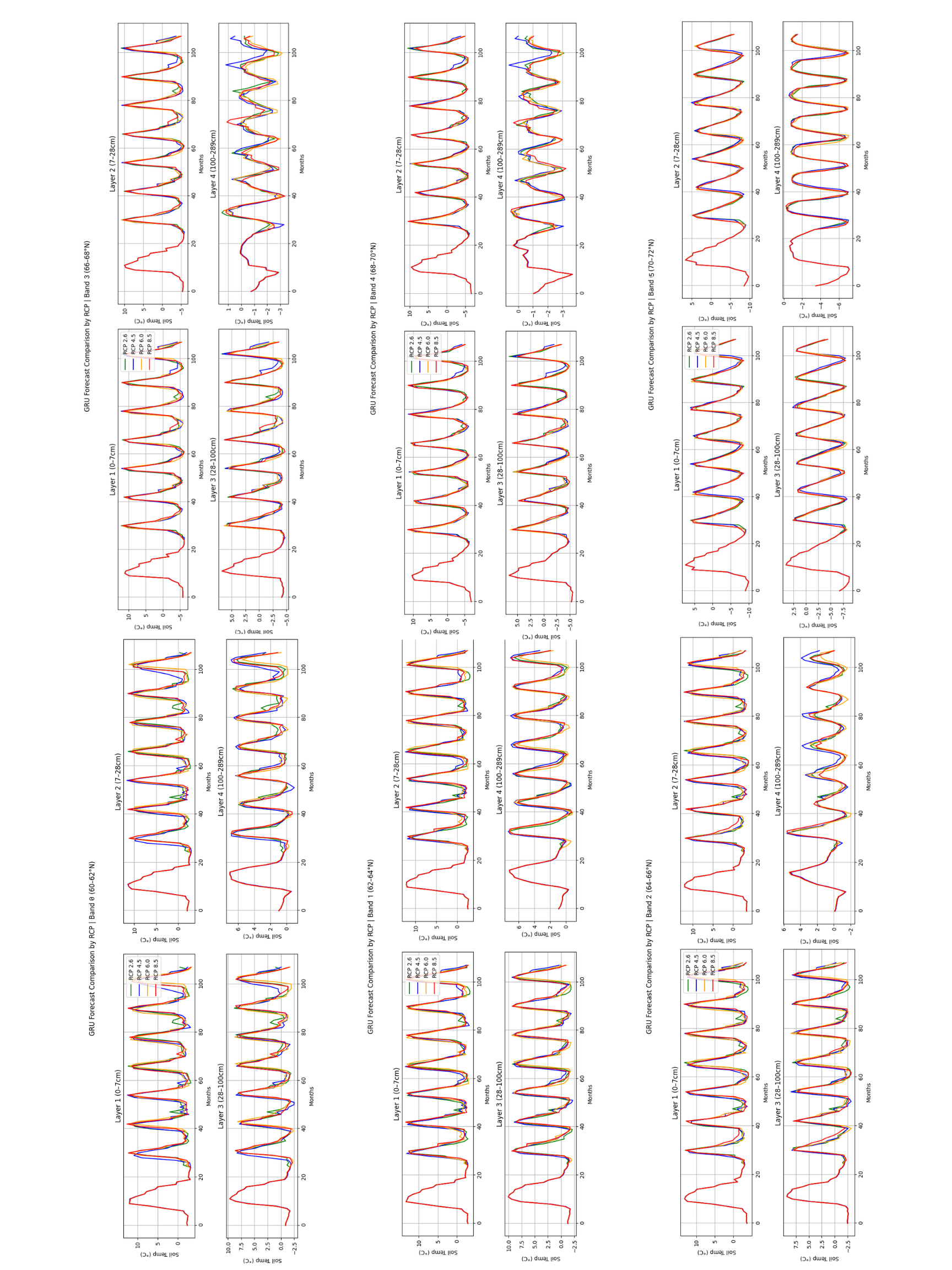}
  \caption{Comparison between GRU predictions per band-layer pair for each RCP scenario. RCP 2.6 is in green, RCP 4.5 is in blue, RCP 6.0 is in yellow, and RCP 8.5 is in red.}
  \label{fig:predictions_gru}
\end{figure}

\begin{figure}[htbp]
  \centering
  \includegraphics[width=0.45\textwidth, angle=-90]{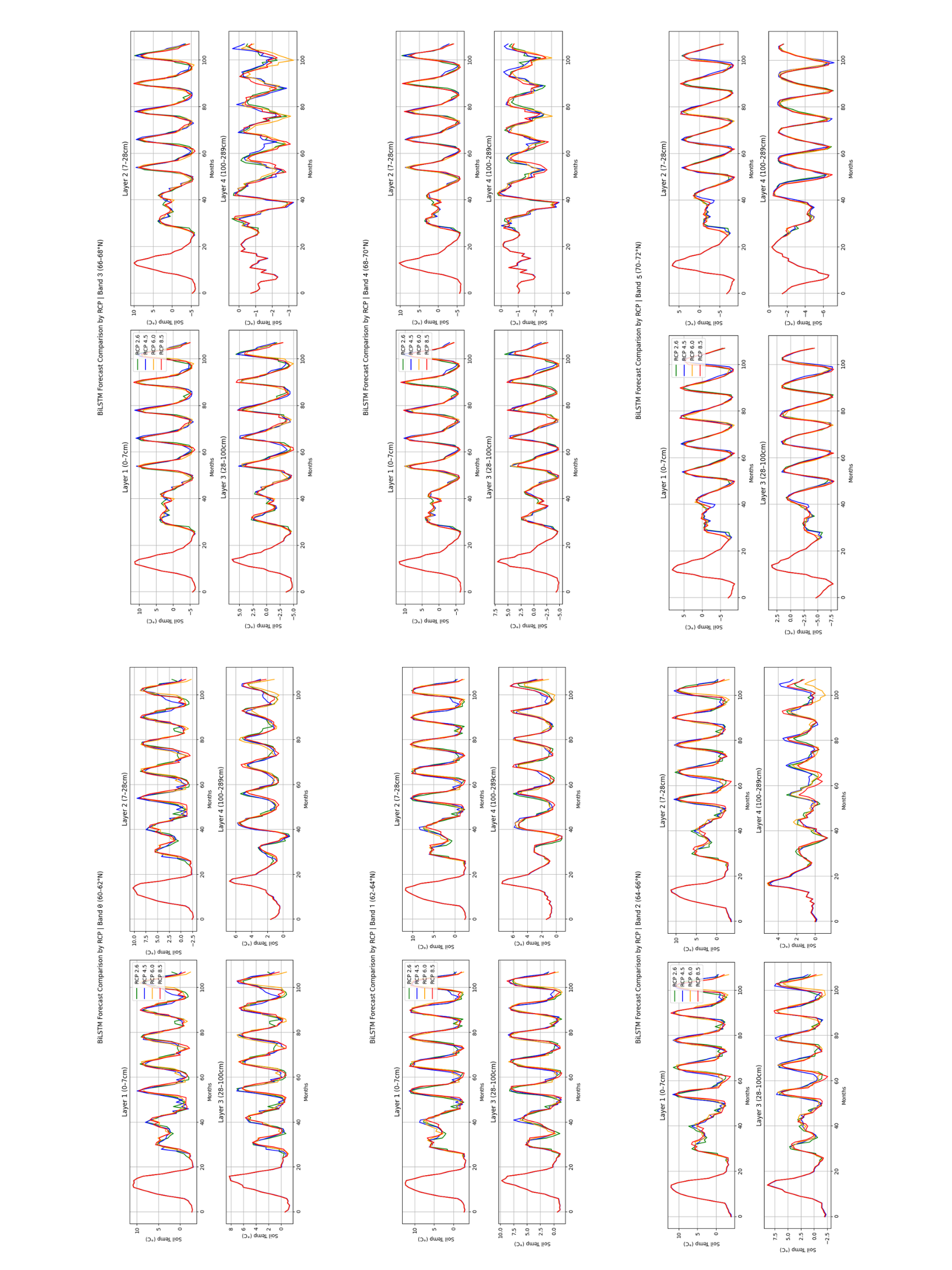}
  \caption{Comparison between BiLSTM predictions per band-layer pair for each RCP scenario. RCP 2.6 is in green, RCP 4.5 is in blue, RCP 6.0 is in yellow, and RCP 8.5 is in red.}
  \label{fig:predictions_bilstm}
\end{figure}

The model ensemble was fed clean, preprocessed data for RCP 2.6, RCP 4.5, RCP 6.0, and RCP 8.5. Temperature stability degrades by increasing order of scenario, with RCP 2.6 being the most optimal scenario, and RCP 8.5 being least optimal. \cref{fig:predictions_tcn}, \cref{fig:predictions_trans}, \cref{fig:predictions_conv1dlstm}, \cref{fig:predictions_gru}, and \cref{fig:predictions_bilstm} display TCN, Transformer, Conv1DLSTM, GRU, and BiLSTM results, respectively per band-layer pair for each pathway. 

Among all the models, the results for each band-layer pair were strikingly similar across RCP scenarios. The lack of divergence suggests that the quantile-mapped IPSL-CM5A-MR inputs were very similar. In fact, the first 20 months of the prediction period perfectly overlap on all 120 graphs, suggesting that the input sequences for those 20 months were identical for all pathways. After the first 20 months, the scenarios shifted slightly; however they followed a similar overall trajectory. This suggests that quantile mapping may have resulted in similar outputs for all RCP scenarios; however, this does not suggest quantile mapping had no benefits. Advantages of quantile mapping will be discussed in \cref{sec:quantile_mapping}.

The networks have also learned periodicity; however they are distributed one period over two years instead of one, not accurately capturing seasonal variations. Nonetheless, capturing general temporal temperature trends, even over a stretched period of time, suggests that the models are promising. The difference in season period may be a result of the intrinsic difference between IPSL-CM5A-MR data and ERA5-Land reanalysis data, specifically for temperature trends and relative positions of inflection points, or it may be a structural consequence of the deep learning models. A strong point in the predictions however, is the understanding of band-wise and depth-wise temperature differences. Similar to ERA5-Land predictions for the year 2023 (\cref{fig:evaluation}), the lower bound of the temperature decreases with increasing band and layer depth. This further implies that band embeddings are feasible and effective, and that spatial embeddings can be of use in future climate deep-learning endeavors. 
 
Convolutional-based models, such as TCNs or Conv1DLSTMS, are particularly strong due to their distinguished ability in detecting localized patterns, such as the monotonic relationship between temperature and band number or layer depth \citep{bai2018, lea2017}. RNNs like the GRU and BiLSTM, on the other hand, may fail to understand the longevity of high-frequency cycle periods due to filtering out noise \citep{hochreiter1997, Chung2014}. The sensitivity of the models may also play a role in near-identical predicted trajectories across scenarios, rendering it unable to process small perturbations and their propagated impact. Previous works have exhibited a similar lack of sensitivity on input configurations; for example, \cite{Ham2019} had noted that the predictive skills of their CNN remained unaffected by dataset perturbations. 

The generally strong performance in accurately predicting soil temperatures from ERA5-Land reinforces the reliability of this proof-of-concept framework.

\subsection{Feature Analysis Results}
\label{sec:feature_analysis}

\begin{figure}[htbp]
  \centering
  \includegraphics[width=0.8\textwidth]{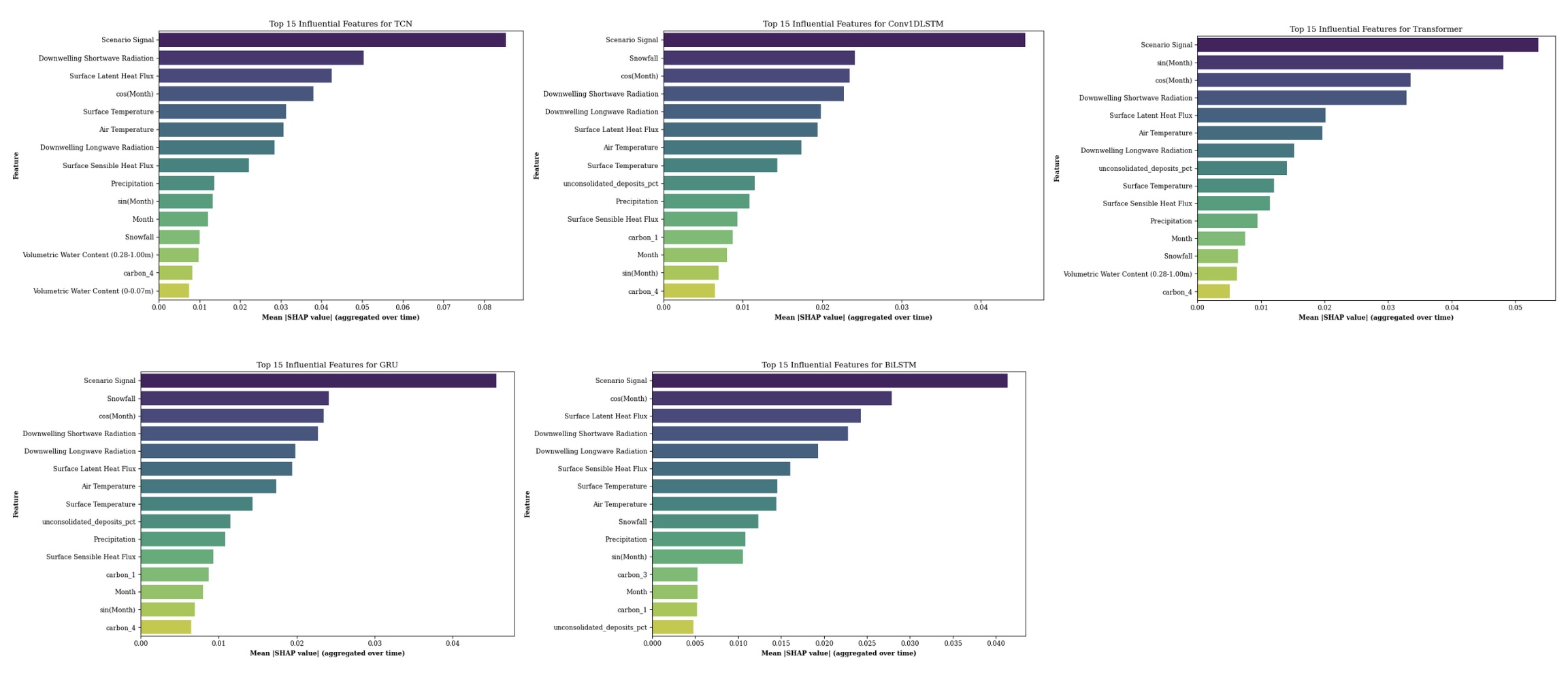}
  \caption{Feature analysis results by model, with bar plots showing mean SHAP values aggregated across all latitude bands and soil layers.}
  \label{fig:shap}
\end{figure}

Feature analysis in this study quantifies the contribution of features for model predictions with computed SHAP scores \cite{Lundberg}. The SHAP scores in this study are decimal values between 0-1; they should not be interpreted as percentages, but rather as absolute contribution values towards model predictions. The top 15 relative feature scores of TCN, Transformer, Conv1DLSTM, GRU, and BiLSTM predictions are plotted in \cref{fig:shap}. 

The unanimously most influential feature was the derived feature, scenario signal ($z$). Due to quantile mapping blurring the differences between scenario data, the significance of scenario signal may have translated less towards RCP divergence as intended, and more towards general temperature periodicity. However, $z$ did increase temperature values as intended, given that it has the highest magnitude among positive SHAP values. The two highest-performing models (TCN and Conv1DLSTM) had snowfall ($P_{SNOW}$) as the second-highest contributing feature. $P_{SNOW}$ would be an appropriate proxy for snow cover, reaffirming previous studies’ emphasis on the importance of snow cover on permafrost insulation \citep{Schuur2008, vanHuissteden2020, gao2022}. Meanwhile, volumetric water content repeatedly had low SHAP scores, suggesting that the models have not learned the zero curtain effect, or that ERA5-Land data did not provide any statistics implying zero curtain effect. Temporal features, such as the sine and cosine of the month index, showed high contribution for all models, particularly for the Transformer model. Transformers, unlike the other four models which were engineered to process data sequentially, do not have a natural sense of temporal order; therefore, they would have heavily relied on positional encoding features to understand temperature cycles. The cosine of the month index was also important in the other models’ predictions as well, but less important for the TCN. Unlike TCNs, RNNs are more prone to blurring cyclic patterns, explaining why they were more reliant on the cosine of the month index. 

\subsection{Impact of Quantile Mapping}
\label{sec:quantile_mapping}

\begin{figure}[htbp]
  \centering
  \includegraphics[width=0.8\textwidth]{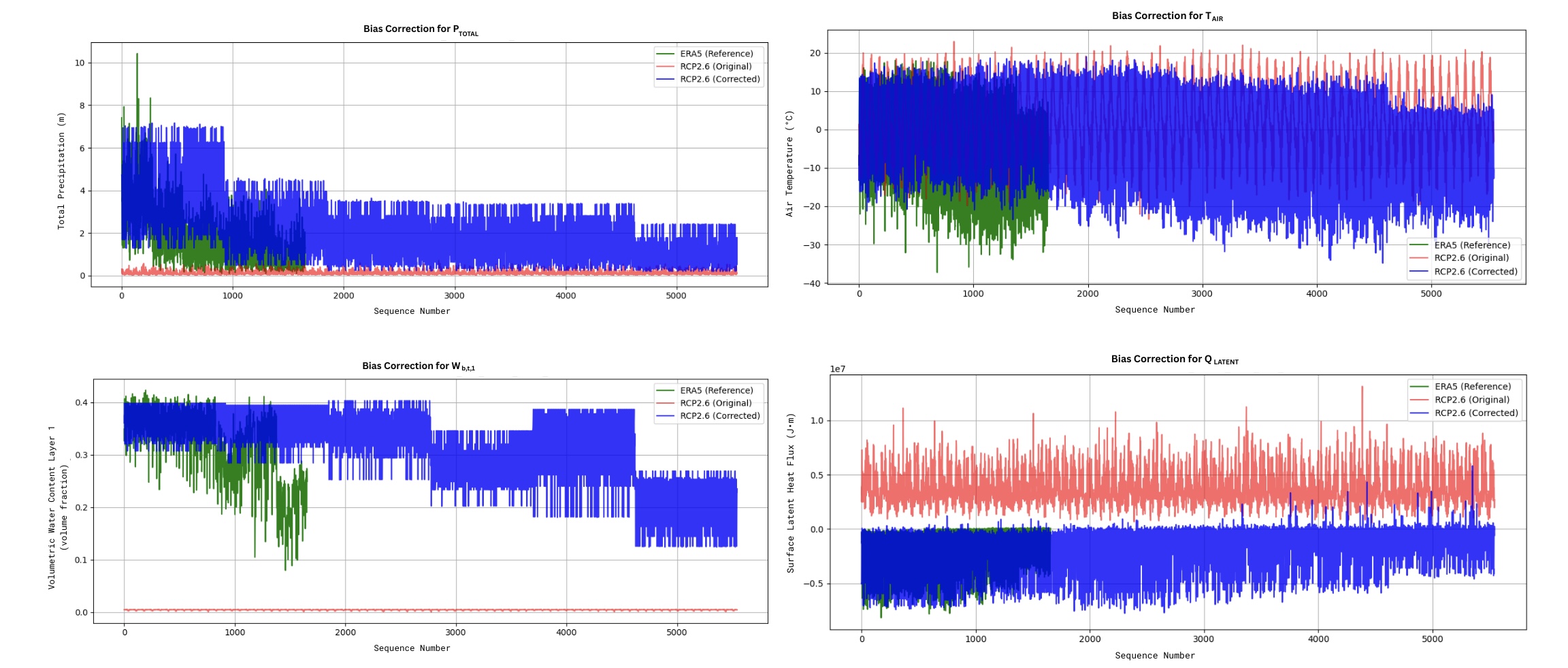}
  \caption{Sequence-wise comparison between historical ERA5-Land data (green), raw RCP 2.6 data (red), and quantile-mapped RCP 2.6 data (blue) for $T_{AIR}$, $Q_{LATENT}$, $P_{TOTAL}$, and $W_{b,t,1}$.}
  \label{fig:bias_correcting}
\end{figure}

Although quantile mapping diluted scenario differences, it also came with significant advantages. Quantile mapping effectively bias-corrected CMIP5 data, addressing systematic differences in spatial and temporal resolution. It redistributed CMIP5 data to have a similar distribution to ERA5-Land data, improving model durability. Redistributed CMIP5 data showed more visual alignment with ERA5-Land than raw ESM simulations (\cref{fig:bias_correcting}). 

\begin{figure}[htbp]
  \centering
  \includegraphics[width=0.6\textwidth]{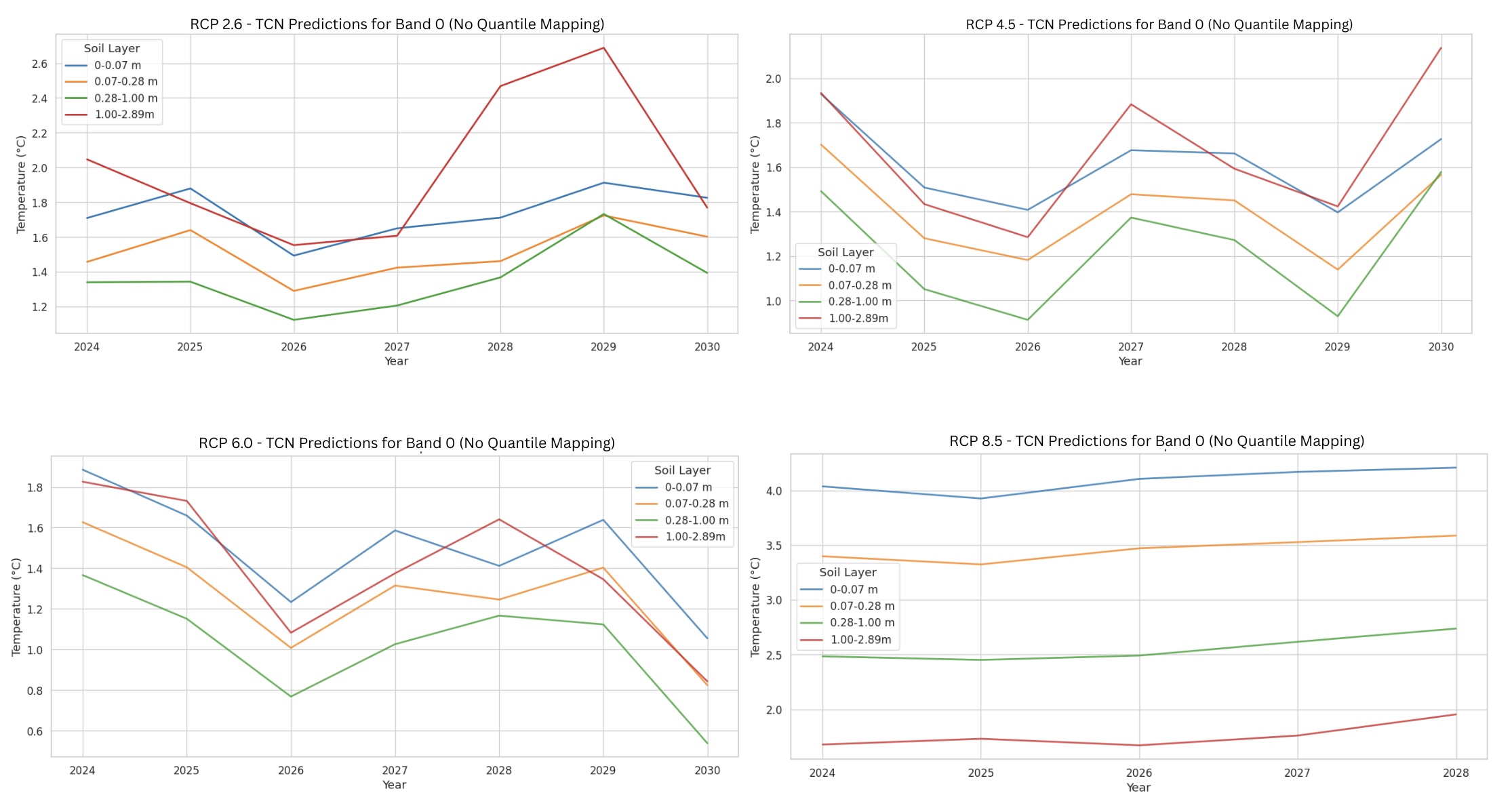}
  \caption{Band 0 TCN Predictions for RCP 2.6 (2024-2030), RCP 4.5 (2024-2030), RCP 6.0 (2024-2030), and RCP 8.5 (2024-2028) per layer.}
  \label{fig:no_quantile}
\end{figure}

In addition, quantile mapping has ensured that models are able to properly observe periodicity in CMIP5 inputs. For example, the TCN, which was the model with the strongest capabilities of identifying seasonal patterns, struggled significantly in predicting sinusoidal trends in raw scenario data devoid of quantile mapping (\cref{fig:no_quantile}). Instead, an entire year of predictions was entirely linear. 

Therefore, quantile mapping was indeed necessary for scenario predictions, otherwise even the most powerful models trained on ERA5-land data would falter for future predictions. 

\section{Conclusion}
This study developed a proof-of-concept framework for permafrost temperature modeling using a sequence-based data pipeline to evaluate the performance of five deep learning models. All models yielded comparable performance results and predictive outcomes, yet relied on different processes. The pipeline integrated band-wise embeddings for spatial sensitivity, sliding-window sequence generation to enable models to learn seasonal context and lag effects, a derived scenario-signal feature to capture long-term climate forcing trends, and quantile mapping to mitigate systemic biases between historical reanalysis data and simulated scenario data. While all models were able to learn seasonal, latitudinal, and depth-based temperature variations with ERA5-Land historical data, their performance on scenario-based predictions was more limited due to effects of quantile mapping and intrinsic biases present in IPSL-CM5A-MR outputs. Nevertheless, quantile mapping proved to be beneficial for enabling recognition of sinusoidal temperature patterns in pathway data. The overall strength of the model suggests that this pipeline may be extended to more robust climate datasets for soil temperature modeling. 

\section{Limitations and Future Work}
The gaps between the ERA5-Land and CMIP5 data archives limited the scope of features for this study. Crucial cryospheric variables in temperature modeling, such as snow cover and snow density (and derived values such as snow thermal conductivity), were not provided by CMIP5, inhibiting the forecast performance of the deep learning models. The IPSL-CM5A-MR model also exhibited limited differences between simulations under different RCP scenarios, causing the models to have similar predictions between scenarios. ERA5-Land reanalysis data also failed to mirror the zero-curtain effect, displaying lower average annual water content with increasing latitude band, preventing the model from learning the zero-curtain effect. 

Future replications of this study would benefit from variables which measure atmospheric carbon and methane content. In addition, a recursive training approach, in which separate models for each scenario are iteratively retrained using training data augmented with each new monthly prediction, offers a promising strategy to mirror compound carbon effects of permafrost thaw. Additionally, concentrating on particular Alaskan boroughs or districts rather than aggregating temperature patterns over a generalized region with wide ranges of elevations may provide a higher resolution of temperature forecast.

% -----------------------------------------------------------------

%\section*{Acknowledgments}
%We would like to acknowledge the assistance of volunteers in putting
%together this example manuscript and supplement.

\bibliographystyle{siamplain}
\bibliography{references}

\end{document}